\documentclass[12pt,aps,prb,preprint]{revtex4}   
\usepackage{amsmath}    
\usepackage{graphicx}   
\begin{document}
\title{On the Relationship between Thermodynamics and Special Relativity}
\author{C. A. Far\'ias}
\affiliation{Departamento de F\'isica, Facultad de Ciencias,
  Universidad de Chile, Casilla 653, Santiago, Chile.} 
\email{cfarias@zeth.ciencias.uchile.cl}
\author{P. S. Moya}
\affiliation{Departamento de F\'isica, Facultad de Ciencias,
  Universidad de Chile, Casilla 653, Santiago, Chile.} 
\email{pmoya@zeth.ciencias.uchile.cl}
\author{V. A. Pinto}
\affiliation{Departamento de F\'isica, Facultad de Ciencias,
  Universidad de Chile, Casilla 653, Santiago, Chile.} 
\email{vpinto@zeth.ciencias.uchile.cl}

\date{\today}

\begin{abstract}
  Starting from a formulation for the $dS$ element that includes movement, and
  considering the variation of the entropy Lorentz invariant, we found the
  relativistic transformations for thermodynamic systems that satisfy
  the three laws of thermodynamics. Particularly, we found the
  temperature and pressure transformations, given by $T'=\gamma T$ and
  $p'=\gamma^2p$ respectively. Furthermore, we show
  that this transformations keeps the form of the state equation for an
  ideal gas in agreement with the relativity principle.  
\end{abstract}


\maketitle

\section{Introduction}\label{intro}

Through history we have seen that many scientific discoveries
have been reached starting from simple ideas. Nowadays
the most part of research is based in the specialization of a single topic.
This fact can be noted in the undergraduate courses in physics where
is not usual to treat recent research themes related to the courses
due to the complexity of those themes. One of the aims of this article
is to show how theoretical research can be done by using simple ideas
that can be learned  in a undergraduated course.

In this article we treat a yet unsolved problem in theoretical
physics: the contruction of a consistent relativistic thermodynamics theory. This
particular problem have generated a long controversy in physics
and can be approached by a student of a thermodynamics course, using only
basic knowledge of thermodynamics and special relativity. Is in this
way that including this particular topic as a part of an
undergraduated thermodynamics or modern physics course could support
the discussion of physics and motivate students to start scientific
research at the undergraduate level.

Let us suppose a system $A$, which is in thermodynamic equilibrium,
and two inertial frames, $I$ and $I'$, where one of them ($I$) is at
rest respective to $A$ and where the other ($I'$) is moving with speed
$\mathbf{w}$ respective to $I$. We orient the axes so the relative motion
of the frames is in the $x$ and $x'$ directions. This is
$\mathbf{w}=w\hat x$. The question is: how the thermodynamics
quantities of $A$ transform between $I$ and $I'$?.  

A century has passed since Einstein, in a first
attempt to answer this question, stated that if the entropy is Lorentz
invariant, and the two first laws of thermodynamics are fulfilled,
then the
temperature transforms as 
\begin{equation}
  \label{traneinstein}
  T'=\frac{1}{\gamma} T\,, 
\end{equation} 
where $\gamma=(1-(w/c))^{-1/2}$ and $c$ is the
speed of light \cite{einstein}; it means that a moving body appears
colder. His temperature transformation was acepted for almost fifty
years, as can be seen in \cite{planck,tolman,pauli, pathria}.  

In 1963, Ott \cite{ott} affirmed that the temperature
transform as 
\begin{equation}
  \label{transott}
  T'=\gamma T 
\end{equation}
by supposing the entropy as a
Lorentz invariant, like other authors did (see for example
\cite{impos} and references therein). This result contradicts the
previous affirmations because in this treatment the body appears hotter.

A few years later, Landsberg \cite{landsvarianza,land1} stated that
temperature should be Lorentz invariant but, years later, he restarted
the problem by saying that it is impossible to obtain a general
transformation of the temperature \cite{impos}. 

Nowadays the question is still open. For example, in \cite{cubero},
we can appreciate an answer derived from the formalism of 
statistical mechanics. Also, even when the topic is, in principle,
theorethical, in this work the authors declare that certains astrophysical systems
might give experimental data which allows us to decide what
transformation is the correct one (see \cite{kania1,kania2} for more
related works).

In this work we approach the problem using the formalism of
thermodynamics. This must be equivalent to an approach from
statistical mechanics since, due the postulate of Gibbs,
thermodynamics quantities are averages of quantities obtained from
statistical mechanics.

We want to emphasize that due the approach used to obtain the results
in this work, it could be a good exercise to undergraduate students in
physics, because basic knowledges in thermodynamics and special
relativity are enough to understand the treatment used here. Also,
the inclusion of this topic in a regular course of thermodynamics can
generate scientific discussion in an actual theme.

\section{Transformation of Thermodynamic Quantities}

Our approach to the problem start by realizing that there is no reason
to suppose that the expression 
\begin{equation}
  \label{dSnorel}
dS=\frac{1}{T}dU+\frac{p}{T}dV-\frac{\mu}{T}dN 
\end{equation}
has the same form for all inertial
frames. Here $T$ is the temperature, $U$ the internal energy, $p$ the
pressure, $V$ the volume, $N$ the number of particles and $\mu$ the
chemical potential. That entropy variation has not the same form in
any inertial frame is because classical thermodynamics has not been 
dedicated to solve the problem about what occurs when there is a
change between reference frames, since it has always considered the
macroscopic properties of matter, but being at rest with respect to it. 

Following a treatment similar to the one of Callen \cite{callen},
we write the $dS$ element like an exact differential but also
adding a new term that involves the relative movement between the
thermodynamic system and the observer. The new term must depends of its
fundamental parameters: the total momentum $\mathbf{P}$ of the thermodynamic system and the relative
speed between the system and an inertial frame. In this case the
velocity is $\mathbf{w}$. 

The term that involves movement comes from the energy of a
relativistic system. If the total mass of our thermodynamical system
is $M$, then due the movement between the system and an inertial
frame, we must add a kinetic energy $E_m$ given by 

\begin{equation}
  \label{E_m}
E_m^2 =
\mathbf{P}^2 + M^2. 
\end{equation}
Then,  
\begin{equation}
dE_m=\frac{\mathbf{P}}{(\mathbf{P}^2 +m^2)^{1/2}}\cdot d\mathbf{P}\, ,
\end{equation}
and therefore

\begin{equation}
  \label{dErel}
dE_m=\mathbf{w}\cdot d\mathbf{P}\, .
\end{equation}

With this we can write the total variation of energy as
\begin{equation}
  \label{dErel2}
  dE=dU+dE_m\, ,
\end{equation}
where $E$ is the total energy of the system.

Later on, using \eqref{dSnorel} and \eqref{dErel}, it follows
\begin{equation}
  \label{dErelrel}
  dE=TdS-pdV+\mu dN+ \mathbf{w}\cdot d\mathbf{P}\, ,
\end{equation}
which gives a relativistic form of writing the element $dS$
for a moving system:
\begin{equation}
  \label{dsrel}
  dS=\frac{1}{T}dE+\frac{p}{T}dV-\frac{\mu}{T}dN-\frac 1T\mathbf{w}\cdot d\mathbf{P}\, .  
\end{equation}

This form of the element $dS$ includes the correction due to the
inclusion of movement between frames. This generalization of
Eq. \eqref{dSnorel} allow us to take into account the relativistic
effects due to the relative motion when analyzing a
thermodynamic system. 

In this way, the entropy of a system must have the
form $S = S(E, V, N, \mathbf{P})$, where the extensive quantities $E$, $V$, $N$
and $\mathbf{P}$ can be measured in any inertial frame. In this
manner, the intensive variables are completely described, not only for
the proper frame of the thermodynamic system, but also for any other
inertial frame. In addition, as can be seen in Eq. \eqref{dsrel}, $dS$ is an exact
differential. Therefore we can define the temperature as
\begin{equation}
  \label{temperatura}
  \frac 1T=\left(\frac{\partial S}{\partial E}\right)_{V,N,\mathbf{P}}\, 
\end{equation}
in any inertial frame, generalizing the well-known expression 
\begin{equation}
  \label{temperaturaor}
  \frac 1T=\left(\frac{\partial S}{\partial U}\right)_{V,N}\, . 
\end{equation}

In order to obtain the transformations of thermodynamic quantities
between $I$ and $I'$, we
first assume that the variation of the entropy $dS$ is Lorentz
invariant, in agreement with all previous works cited here.
This is, 
\begin{equation}
  \label{dS=dS'}
  dS=dS'\,
\end{equation}
where $S$ and $S'$ are the entropies measured in the $I$ and $I'$
frames respectively.
For the other quantities we have that the number of particles $N$ is Lorentz
invariant, and in this case, between $I$ and $I'$, the volume $V$ and
the energy $E$ transforms as
\begin{eqnarray}
V'&=&\frac 1\gamma V\, ,\label{transvolume}\\
E'&=&\gamma E\, , \label{transenergy}
\end{eqnarray}
respectively, where $E'$ and $V'$ are measured in $I'$.
 
As the differential form for the entropy guarantees
Eq. \eqref{temperatura}, using Eq. \eqref{dS=dS'} and Eq. \eqref{transenergy}, it follows
that the temperatute $T'$, measured in the $I'$ frame, is related to
$T$ by
\begin{equation}
\label{transT}
\frac{1}{T'}=\left(\frac{\partial S'}{\partial
    E'}\right)_{V',N',\mathbf{P}'}=\frac{1}{\gamma}\left(\frac{\partial S}{\partial
    E}\right)_{V,N,\mathbf{P}}=\frac{1}{\gamma T}\, ,
\end{equation}
from where the temperature transformation
\begin{equation}
T'=\gamma T\, , \label{temptrans}
\end{equation}
is obtained. This transformation is in agreement with the one by Ott
\cite{ott} and other authors \cite{arzelies, gamba, sutcliffe}. Also Eq. \eqref{temptrans} is in
agreement with the three laws of thermodynamics, unlike what was
proposed by Einstein (see Eq \eqref{traneinstein}).  

For the pressure $p'$, using Eq. \eqref{dsrel}, we have
\begin{equation}
  \label{p/T}
  \frac pT = \left(\frac{\partial S}{\partial V}\right)_{E,\mathbf{P},N}\,,
\end{equation}
Then,  using Eq. \eqref{dS=dS'}, \eqref{transvolume}, \eqref{temptrans} and
\eqref{p/T} we obtain
\begin{equation}
  \label{prest}
  p'=T'\left(\frac{\partial S'}{\partial V'}\right)_{E',\mathbf{P}',N'}=\gamma T\left(\frac{\partial S}{\partial
    (V/\gamma)}\right)_{E,\mathbf{P},N} \, .
\end{equation}
This is, the pressure transforms as
\begin{equation}
p'=\gamma^2p\, .  \label{presstrans}
\end{equation}
This result, previously obtained by Sutcliffe \cite{sutcliffe}, is
natural when we accept that Eq. \eqref{p/T} is the correct definition
of pressure in any inertial frame.

Following the same argument, for the $\mu'$ potential,
\begin{equation}
\label{mutrans}
\mu'=\gamma\mu\, ,
\end{equation}
it is obtained. If we note that $\mu$ is an energy, and energy transforms
as $\gamma$, then, this $\mu$ transformation was expected.

As a example, from Eq. \eqref{transvolume}, Eq. \eqref{temptrans} and
Eq. \eqref{presstrans}, it follows that
\begin{equation}
  \label{idealgas}
  \frac{pV}{Nk_BT}=\frac{p'V'}{N'k_BT'}\, ,
\end{equation}
which means that the equation of state of an ideal gas is Lorentz
invariant. This means that, for any inertial observer, an ideal gas
is still an ideal gas, regardless of relative inertial motion.

\section{Final discussion}

The development of relativistic thermodynamics has been a complex and
hard to treat theme for a long time. There has been many 
proposals, many of them contradict each other. In this article we have
formulated a theory of thermodynamics where the movement 
between inertial frames is considered. This led us to find a new
expression for the entropy by considering the kinetic energy due
to the relative motion between frames. The new entropy $S=S(E,V,\mathbf{P},N)$ now
depends on the total energy $E$ instead of the internal energy
$U$ and also of the momentum of the system $\mathbf{P}$ measured in
a inertial frame. This choice allowed us to write the $dS$ element as an
exact differential that now has a new term,
$(1/T)\mathbf{w}\cdot d\mathbf{P}$, which is related to the relative
movement between frames. Using this new expression for $dS$, just as
it is done in classical thermodynamics,
we were able to define intensive thermodynamic quantities by taking
partial derivatives of $S$ respect to the extensive quantities. This
formalism satisfies the three laws of thermodynamics and becomes 
the usual one in the limit $\mathbf{w}\to0$.

Using this formalism we obtained the transformations given by Eq. \eqref{temptrans},
\eqref{presstrans} and \eqref{mutrans} for the temperature, the
pressure and the chemical potencial
respectively. These transformations are completely general for any
system that satisfy the three laws of thermodynamics.

Taking this last argument we were able to state why the temperature
transformation given by Eq. \eqref{temptrans} is the correct one in a
relativistic thermodynamic theory. In thermodynamics the temperature is that
quantity which, at equilibrium, is given by
Eq. \eqref{temperatura}. Therefore, using that $dS$ is an exact
differential and the known transformations of the extensive
quantities, $E,V$ and $N$, it follows that the transformation of temperature which is in agreement
with the three laws of thermodynamics is the one given by Eq. \eqref{temptrans}.

In addition, as a natural consecuence of our treatment, we found that the
pressure transformation is given by Eq. \eqref{presstrans}. This
transformation does not agree with the one stated by
many authors, which consider pressure to be  Lorentz invariant. Nevertheless
the transformation given by Eq. \eqref{presstrans} must be the correct one in a
relativistic thermodynamics theory because it corresponds to the
definition of pressure given by thermodynamics, as is shown in
Eq. \eqref{p/T}.

Finally we showed that, if we accept our transformations for the
pressure, temperature, number of particles and volume, the equation of state of an ideal gas is Lorentz
invariant, which is in accord with the first
principle of special relativity. It should be noted that if we have
taken the pressure and temperature as Lorentz invariants the equation
of state of an ideal gas would not be satisfied. Also, if we had
considered the pressure as Lorentz invariant and the
transformation of temperature as $T'=\gamma^{-1}T$ the equation of state of
an ideal gas would have been satisfied but not the third law of thermodynamics.

\begin{acknowledgments}
We would like to thank Dr. G. Guti\'errez for his motivation to do
this work as an extension of his Thermodynamics lessons.  We also
would like to thanks Dr. J. Zanelli and Dr. R. Tabensky for their 
disposition and useful discussions and advise while doing this
work. P. S. Moya and V. A. Pinto are grateful to CONICyT Doctoral Fellowship.
C. A. Far\'\i as is grateful to CONICyT Master Fellowship.
\end{acknowledgments}

\end{document}